\documentclass[conference,a4paper]{IEEEtran}
\usepackage{cite}
\usepackage{graphicx,color,epsfig,rotating}
\usepackage{amsfonts,amsmath,amssymb,bbm}
\usepackage{algorithm}
\usepackage{algpseudocode}
\usepackage{subfigure}
\usepackage{amsmath}
\usepackage{cite}
\usepackage{placeins}
\usepackage{graphicx}
\usepackage[latin1]{inputenc}
\usepackage{amssymb}
\usepackage{multirow}
\usepackage{stfloats}
\usepackage{tabularx} 
\usepackage{booktabs} 
\usepackage{url}
\usepackage{bm}
\usepackage{soul}
\usepackage{float}
\usepackage{lipsum}     
\usepackage{cuted} 
\usepackage{pstricks}

\long\def\comment#1{}

\newfont{\bbb}{msbm10 scaled 700}

\newfont{\bb}{msbm10 scaled 1100}

\newcommand{\uv}{{\bf u}}

\newcommand{\vv}{{\bf v}}

\newcommand{\yv}{{\bf y}}

\newcommand{\be}{\begin{equation}}
\newcommand{\ee}{\end{equation}}
\newcommand{\bea}{\begin{eqnarray}}
\newcommand{\eea}{\end{eqnarray}}

\newtheorem{example}{Example}
\newtheorem{theorem}{Theorem}

\newtheorem{corollary}{Corollary}

\usepackage{multicol}
\makeatletter
\newcommand{\subalign}[1]{
  \vcenter{%
    \Let@ \restore@math@cr \default@tag
    \baselineskip\fontdimen10 \scriptfont\tw@
    \advance\baselineskip\fontdimen12 \scriptfont\tw@
    \lineskip\thr@@\fontdimen8 \scriptfont\thr@@
    \lineskiplimit\lineskip
    \ialign{\hfil$\m@th\scriptstyle##$&$\m@th\scriptstyle{}##$\crcr
      #1\crcr
    }%
  }
}
\makeatother

\ifodd 1

\else

\fi

\ifodd 1

\else

\fi

\begin{document}
\title{A New Design of Cache-aided Multiuser Private Information Retrieval with Uncoded Prefetching}
\author{\IEEEauthorblockN{Xiang Zhang\IEEEauthorrefmark{1}, Kai Wan\IEEEauthorrefmark{2}, 
Hua Sun\IEEEauthorrefmark{3}, Mingyue Ji\IEEEauthorrefmark{1} and Giuseppe Caire\IEEEauthorrefmark{2} }
\IEEEauthorblockA{Department of Electrical and Computer Engineering, University of Utah\IEEEauthorrefmark{1}\\
Department of Electrical Engineering and Computer Science, Technische Universit\"at Berlin\IEEEauthorrefmark{2}\\
Department of Electrical Engineering, University of North Texas\IEEEauthorrefmark{3}\\
Email: \IEEEauthorrefmark{1}\{xiang.zhang, mingyue.ji\}@utah.edu, \IEEEauthorrefmark{2}\{kai.wan, caire\}@tu-berlin.de, 
\IEEEauthorrefmark{3}hua.sun@unt.edu}}

\maketitle

\begin{abstract} 
In the problem of cache-aided multiuser private information retrieval (MuPIR), a set of $K_{\rm u}$ cache-equipped  users wish  to privately download a set of messages from $N$ distributed databases each holding a library of $K$ messages. The system works in two phases: the {\it cache placement (prefetching) phase} in which the users fill up their cache memory, and the {\it private delivery phase} in which the users' demands are revealed and they download an answer from each database so that the their desired messages can be recovered while each individual database learns nothing about the identities of the requested messages. The goal is to design the placement and the private delivery phases such that the \emph{load}, which is defined as the total number of downloaded bits normalized by the message size, is minimized given any user memory size. This paper considers the MuPIR problem with two messages, arbitrary number of users and databases where uncoded prefetching is assumed, i.e., the users directly copy some  bits from the library as their cached contents. We propose a novel MuPIR scheme inspired by the Maddah-Ali and Niesen (MAN) coded caching scheme. The proposed scheme achieves lower load than any existing schemes, especially the product design (PD), and is shown to be optimal within a factor of $8$ in general and exactly optimal at very high or very low memory regimes.

\end{abstract}
\section{Introduction} 
The problem of private information retrieval (PIR)\cite{chor1995private} seeks the most efficient way for a user to retrieve one of the $K$ messages in the library from $N$ distributed databases while keeping the information identity  secret from each database. Recently, Sun and Jafar (SJ) characterized the capacity of the PIR problem as $\left (1+\frac{1}{N}+\cdots+\frac{1}{N^{K-1}}   \right)^{-1}$\cite{sun2017capacity}.  Based on the achievability and converse techniques developed in \cite{sun2017capacity},  many variants of the PIR problem, including colluding databases, storage-constrained databases and symmetric PIR, have been studied. The effect of storage (i.e., cache or side information) is brought into consideration in the cache-aided PIR problem which has gained significant attention recently. Two different privacy models are commonly considered. In one line of research \cite{tandon2017capacity,wei2018fundamental,8362308}, the user-against-database privacy model is used where individual databases are prevented from learning the single-user's demand. \cite{tandon2017capacity} showed that the load $\left (1-\frac{M}{K} \right)\left (1+\frac{1}{N}+\cdots+\frac{1}{N^{K-1}}   \right)$ achieved by a simple memory sharing scheme is optimal for any  user cache size  $M\in[0,K]$ if the databases know the cache of the user, implying that globally known side information does not help to improve retrieval efficiency. However, if the databases are unaware of the user's   cache, there is an ``unawareness gain" in capacity shown by  \cite{wei2018fundamental,8362308}. 
Another line of research \cite{wan2019coded,kamath2019demand, sarvepalli2019subpacketization,wan2019device} considers the user-against-user privacy model where users are prevented from learning each other's demands. The authors in \cite{wan2019coded} introduced the problem of coded caching with private demands in which the users of a shared-link caching network are prevented from knowing each others' demands. Order optimal schemes were proposed based on the use of virtual users. The subpacketizaiton issue of demand-private coded caching was studied in  \cite{sarvepalli2019subpacketization}. \cite{9330765} considered the same setting but with colluding users. Later, coded caching with private demands was extended to the device-to-device (D2D) caching networks \cite{wan2019device}. In general, the exact capacity characterization remains open for these problems.

Introduced in \cite{9174083}, the problem of \emph{cache-aided multiuser PIR} (MuPIR) considers a basic setting including $K_{\rm u}$  cache-equipped users whose cached contents are known by the $N$ distributed databases. The goal of this problem is to design the cache placement and private delivery phases such that the download from the databases is minimal while preserving the privacy of the users' demands from each database.  An order optimal scheme, referred to as \emph{product design}, was proposed in  \cite{9174083}. It uses the cache placement proposed by Maddah-Ali and Niesen (MAN) for the original shared-link coded caching problem~\cite{maddah2014fundamental} to benefit from the multicast gain provided by coded caching and embedded the SJ PIR code \cite{sun2017capacity} to preserve user demand privacy. In \cite{9155254}, an achievable scheme based on a novel  cache-aided interference alignment (CIA) idea was proposed for the MuPIR problem with two messages, two users and arbitrary number of databases, which is shown to be optimal when $N \in \{2,3\}$. 

In this paper, for the cache-aided MuPIR problem, we propose an improved achievable scheme with a novel uncoded prefetching for the case of two messages, arbitrary number of users and databases.
The proposed scheme achieves the optimal memory-load trade-off for the case of $K=K_{\rm u}=2$ and  $N=2,3$ which is consistent with the optimality results of \cite{9155254}. The proposed scheme has strictly better performance than the product design and is order optimal within a factor of $8$. Moreover, the proposed scheme requires a smaller message size (linear in $N$) than the product design which requires a message size quadratic in $N$  when $K=2$. 
In addition, the proposed scheme attains the optimal load in the following two cases: \emph{1)} When $K=2$, $K_{\rm u}=3$, and $N\ge 2$, the proposed scheme  achieves the optimal load (under uncoded prefetching) in the small memory regime; \emph{2)} When $K=2$, $K_{\rm u}\ge 2$, and $N\ge 2$, the proposed scheme achieves the optimal load when the memory size is at least $\frac{2(K_{\rm u}-1)(N-1)}{K_{\rm u}(N-1)+1}$.

\paragraph*{Notation convention}
$\mathbb{Z}^{+}$ denotes the set of positive integers. For some $n\in\mathbb{Z}^+, [n]\triangleq \{1,2,\ldots,n\}$. $\binom{[n]}{m}$ denotes a set containing all the subsets of $[n]$ of size $m$, i.e., $\binom{[n]}{m}\triangleq \left\{ \mathcal{A}\subseteq [n]: |\mathcal{A}|=m  \right  \}$. For an index set $\mathcal{I}=\{i_1,\ldots,i_n\}$, $A_{\mathcal{I}}$ denotes the set $A_{\mathcal{I}} \triangleq \{A_{i_1},\ldots,A_{i_1}\}$. We write $A_{[m:n]}$ as $A_{m:n}$ for  short. For an index vector ${I}=(i_1,\ldots,i_n)$, $A_{{I}}$ denotes the vector $A_{{I}}\triangleq (A_{i_1},\ldots,A_{i_n})$. $\bm{0}_n \triangleq (0,0,\ldots,0)$ and $\bm{1}_n \triangleq (1,1,\ldots,1)$ (of length $n$). $\mathbf{I}_n$ denotes the $n\times n$ identity matrix.  The matrix $[a;b]$ is written in the MATLAB form, representing $[a,b]^{\text{T}}$.

\section{Problem Formulation}
We consider the $(K,K_{\rm u},N)$ cache-aided MuPIR problem including $N\ge 2$ distributed (non-colluding) databases (DBs) with access to a library of $K$ messages denoted by $W_1,\ldots,W_K$ each of which is uniformly distributed over $[2^{L}]$, and $K_{\rm u}$ users each equipped with a cache memory of   $ML$ bits. Let $Z_1,\ldots,Z_{K_{\rm u}}$ denote the caches of the users. The system operates in two phases, i.e., a cache placement (or prefetching) phase followed by a private delivery phase. In the prefetching phase, the users fill up their caches   without the knowledge of their future demands. We consider uncoded prefetching where each user directly copies a subset of the bits   in the library. We also assume that $Z_1,\ldots,Z_{K_{\rm u}}$ are known by each DB. In the private delivery phase, each user $u\in[K_{\rm u}]$ wishes to retrieve a message $W_{\theta_u}$ from the DBs. Let the demand vector $\bm{\theta}\triangleq (\theta_1,\ldots,\theta_{K_{\rm u}})\in [K]^{K_{\rm u}}$ represent the demands of the users.
Depending on the caches $Z_1,\ldots,Z_{K_{\rm u}}$ and the demand vector $\bm{\theta}$, the users cooperatively generate $N$ queries $Q_1^{[\bm{\theta}]},\ldots,Q_N^{[\bm{\theta}]}$ and then send $Q_n^{[\bm{\theta}]}$ to DB $n$. Upon receiving the query, DB $n$ responds with an answer $A_n^{[\bm{\theta}]}$ broadcasted to all users. $A_n^{[\bm{\theta}]}$ is assumed to be a deterministic function of
the query  $Q_n^{[\bm{\theta}]}$ and the messages $W_{1:K}$.
%
Each user should recover its desired message with the answers downloaded from the DBs, which can be expressed as 
\be \label{eqn: decodabilty}
H\left(W_{\theta_u}|{Q^{[\bm{\theta}]}_{1:N}},{A^{[\bm{\theta}]}_{1:N}},Z_u\right)=0,\quad \forall u\in[K_{\rm u}].
\ee
User demand privacy requires that from each individual DB's perspective,  $\bm{\theta}$ should be independent of the information available to that DB, which can be expressed in terms of mutual information as
\be \label{eqn: privacy}
I\left(\bm{\theta};{ Q^{[\bm{\theta]}}_{n}, A^{[\bm{\theta}]}_{n}},W_{1:K},Z_{1:K_{\rm u}}\right)=0,\quad \forall n\in[N].
\ee

The \emph{load} of the MuPIR problem, denoted by $R$, is defined as the average  number of bits downloaded from the DBs per useful message bit. Let $D$ denote the total number of bits broadcasted from the DBs, then 
\be 
R\triangleq  \frac{D}{L}={\sum_{n=1}^{N}H ({ A^{[\bm{\theta}]}_{n})}}/{L}.
\ee
A memory-load pair $(M,R)$ is said to be achievable if there exists a MuPIR scheme satisfying the decodability constraint (\ref{eqn: decodabilty}) and the privacy constraint (\ref{eqn: privacy}). The goal is to design the cache placement and the private delivery phases such that the load $R$ can be minimized given  any memory size $M$.

\section{Main Result}
\begin{theorem} 
\label{thm 1}
For the cache-aided MuPIR problem with $K=2$ messages, $K_{\rm u}\ge 2$ users and $N\ge 2$ databases, the following memory-load pairs $(M_t,R_t),\forall t\in[K_{\rm u}-1]$ are achievable:  
\begin{subequations}
\label{eq: thm 1 pairs}
\begin{align}
M_t&=\frac{K\binom{K_{\rm u}-1}{t-1}(N-1)   }{\binom{K_{\rm u}}{t}(N-1)+\binom{K_{\rm u}}{t+1}},\label{eq: thm 1 memory} \\
R_t &=\min\left\{\frac{\binom{K_{\rm u}}{t+1}(N+1)}{\binom{K_{\rm u}}{t}(N-1)+\binom{K_{\rm u}}{t+1}},K-M_t\right\}\label{eq: thm 1 load}. 
\end{align}
\end{subequations}
For $M\ne M_t$, the lower convex envelope of the pairs $(M_t,R_t),\forall t\in[K_{\rm u}-1]$, $(0,2)$ and $(2,0)$ is achievable. 
\hfill $\square$
\end{theorem}
\begin{IEEEproof}
The memory-load pairs $(0,2)$ and $(2,0)$ are trivially achievable. The second term $K-M_t$ in (\ref{eq: thm 1 load}) can be achieved by  a naive scheme where we let each user store the same $M_t/K$ fraction of each message and then let the databases broadcast the remaining $1-M_t/K$ fraction of each message, achieving the load $K(1-M_t/K)=K-M_t$. In Section \ref{section: proof of thm 1}, we propose a new uncoded scheme 
to achieve the memory-load pairs $(M_t,R_t),\forall t\in[K_{\rm u}-1]$. 
\end{IEEEproof}

\begin{corollary} 
\label{corollary: comparison to PD}
For any $K=2, K_{\rm u}\ge 2,N\ge 2$ and $M\in[0,2]$, the achievable load in Theorem \ref{thm 1} is no larger than the load achieved by the product design \cite{9174083}.   Moreover, the load in Theorem \ref{thm 1} is optimal within a factor of $8$.
\hfill $\square$
\end{corollary} 

\begin{IEEEproof}
See Appendix \ref{appendix: proof Corollary 1}. The order optimality directly comes from the order optimality of the product design.
\end{IEEEproof}

\begin{figure}
\centering
\includegraphics[width=0.44\textwidth]{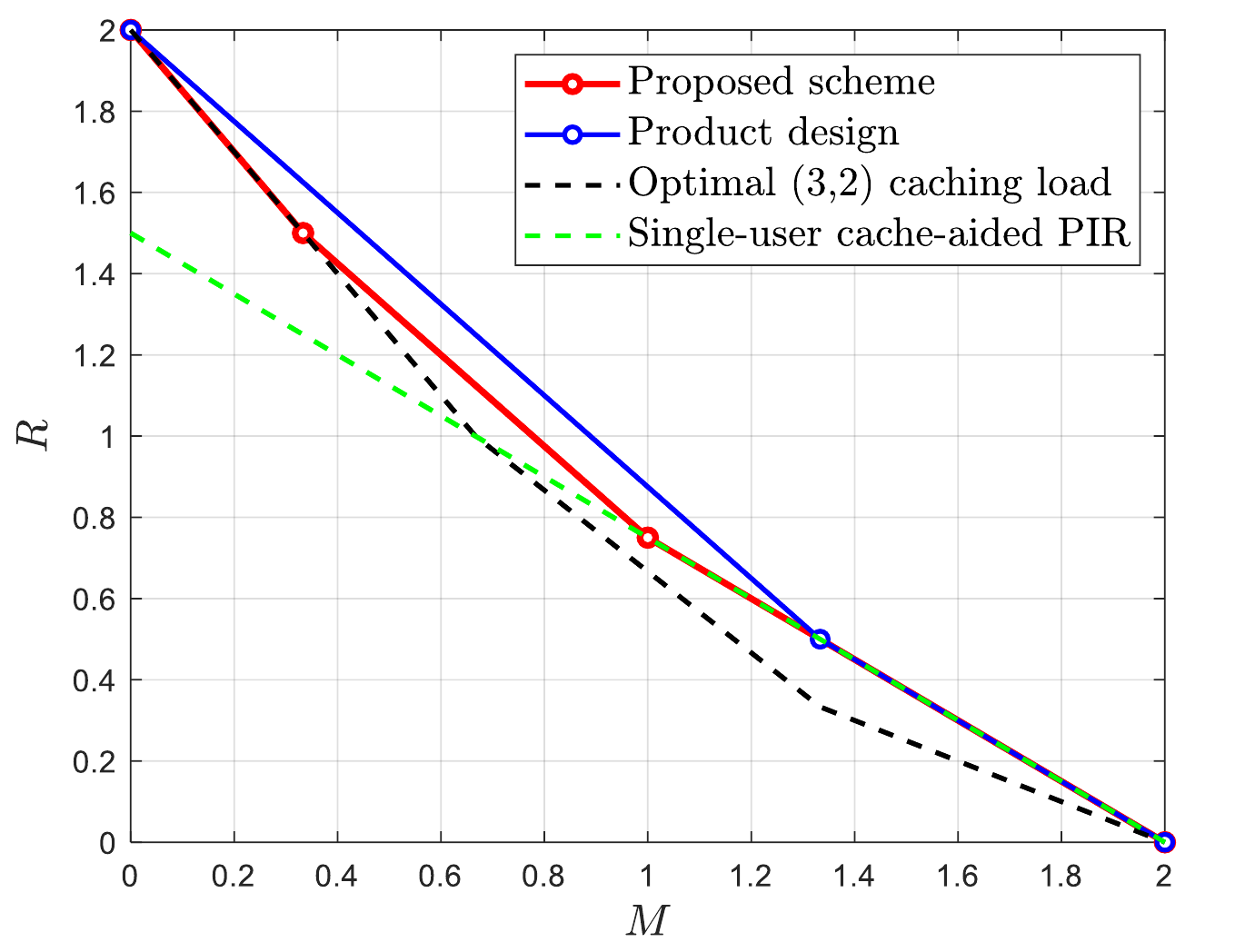}
\caption{The achievable load of the proposed scheme and the product design for the $(K,K_{\rm u},N)=(2,3,2)$ cache-aided MuPIR problem.
}
\label{fig2}
\end{figure}

Fig.~\ref{fig2} compares the achievable load in Theorem \ref{thm 1} to that of the product design for $(K,K_{\rm u},N)=(2,3,2)$. It can be seen that the proposed scheme outperforms the product design. When $M\le 1/3$, the achievable load coincides with the optimal  load of the shared-link coded caching problem~\cite{maddah2014fundamental}, implying that imposing  privacy does not incur any extra download in the small memory regime. When $M\ge 1$, the achievable load coincides with the optimal single-user cache-aided PIR load \cite{tandon2017capacity}, implying that adding more users incurs no extra download when $M$ is large.

\begin{corollary} (\textbf{Optimality in small memory regime})
\label{corollary: optimality at small memory regime}
For $K=2,N\ge 2$,  $K_{\rm u}=2$ and $3$, the achievable load in Theorem \ref{thm 1}  is optimal when $M\le M_1=\frac{2(N-1)}{K_{\rm u}(N-1)+K_{\rm u}(K_{\rm u}-1)/2}$ under the constraint of uncoded prefetching. \hfill $\square$
\end{corollary}
\begin{IEEEproof}
According to \cite{8226776}, for $K_{\rm u}\in\{2,3\}$, the optimal $(K_{\rm u },2)$ caching load is $R^{\star}_{\rm caching}(M)=2-\frac{2-r}{m}M$ when $M\le m$ where $m = 2/K_{\rm u}, r= (K_{\rm u}-1)/2$. For $t=1$, we have $(M_1,R_1)=\left( \frac{m(N-1)}{r+N-1} ,\frac{r(N+1)}{r+N-1}  \right)$, which lies on $R^{\star}_{\rm caching}(M)$, so the load in Theorem \ref{thm 1} is optimal when $M\le M_1$.
\end{IEEEproof}

\begin{corollary} (\textbf{Optimality in large memory regime})\label{corollary: optimality at large memory regime}
For $K=2, K_{\rm u}\ge 2$ and $N\ge 2$, the achievable load in Theorem \ref{thm 1} is optimal when $M\ge M_{K_{\rm u}-1}=\frac{2(K_{\rm u}-1)(N-1)}{K_{\rm u}(N-1)+1}$. 
\hfill $\square$
\end{corollary}
 \begin{IEEEproof}
 The memory-load pair of Theorem \ref{thm 1} corresponding to $t=K_{\rm u}-1$ is $M_{K_{\rm u}-1}= \frac{2(K_{\rm u}-1)(N-1)}{K_{\rm u}(N-1)+1}$, $R_{K_{\rm u}-1}=\frac{N+1}{K_{\rm u}(N-1)+1} $.
 It can be seen that $(M_{K_{\rm u}-1}, R_{K_{\rm u}-1})$ lies on the single-user cache-aided PIR converse bound $R^{\rm SU}(M)=(1-{M}/{2})(1+{1}/{N})$ \cite{tandon2017capacity}. Also, by memory sharing between $(M_{K_{\rm u}-1}, R_{K_{\rm u}-1})$ and $(2,0)$, the load $R=(1-{M}/{2})(1+{1}/{N})$ is achievable. Since increasing the number of users while preserving demand privacy can only possibly increase the load, the achievable load in Theorem \ref{thm 1} is optimal when $M\ge M_{K_{\rm u}-1}$. 
 \end{IEEEproof}

\section{Examples}
In this section, we provide  two examples to highlight the main idea of the proposed scheme for the $(2,3,2)$ MuPIR problem. Let $A,B$ denote the two messages. We show the achievability of the memory-load pairs $(1/3,3/2)$ and $(1,3/4)$ in the following.
\begin{example} 
\label{example (1/3,3/2)}
(\textbf{Achievability of $(1/3,3/2)$}) Consider the cache-aided MuPIR problem with $(K,K_{\rm u},N)=(2,3,2)$ and  the memory-load pair corresponding to $t=1$ in~\eqref{eq: thm 1 pairs}, which is $(1/3,3/2)$.

\emph{1) Cache placement:} Suppose each message consists of $L=6$ bits, i.e., $A=(A_1,\ldots,A_6),B=(B_1,\ldots,B_6)$. Each user stores one bit of each message, i.e., the cache placement is $Z_1=\{A_1,B_1\}, Z_2=\{A_2,B_2\}$ and $Z_3=\{A_3,B_3\}$.

\emph{2) Private delivery:} Suppose $\bm{\theta}=(1,1,2)$.  This phase consists of three steps each corresponding to a subset $\mathcal{S}$ of $t+1=2$ users. Therefore, the answer of 
of DB $n$ consists of three parts $A_n^{[\bm{\theta}]}=\left( A_{n,\mathcal{S}}^{[\bm{\theta}]}:\forall \mathcal{S}\in \binom{[3]}{2} \right)$. For each $\mathcal{S}$, the delivery involves a specific set of message bits denoted by $\mathcal{M}_{\mathcal{S}}$. The delivery scheme is described as follows.

For $\mathcal{S}_1=\{1,2\}$, the involved message bits are $\mathcal{M}_{\{1,2\}} =\{A_1,A_2,A_4,B_1,B_2,B_4\}$. The answer of DB 1 (for $\mathcal{S}_1$)  contains one linear combination of the message bits in $\mathcal{M}_{\{1,2\}}$. The answer of DB 2 (for $\mathcal{S}_1$) consists of two linear combinations of the bits in $\mathcal{M}_{\{1,2\}}$, i.e., $A_{2,\mathcal{S}_1}^{[\bm{\theta}]}=(A_{2,\mathcal{S}_1,1}^{[\bm{\theta}]}   , A_{2,\mathcal{S}_1,2}^{[\bm{\theta}]}   )$. The answers can be written as
\begin{equation}
\label{eq: answer for S1}
\begin{bmatrix}
A_{1,\mathcal{S}_1}^{[\bm{\theta}]}\\
A_{2,\mathcal{S}_1,1}^{[\bm{\theta}]}  \\
A_{2,\mathcal{S}_1,2}^{[\bm{\theta}]}  
\end{bmatrix}= 
\setlength\arraycolsep{3.7pt}
\begin{bmatrix}
u_{1,1}^{\mathcal{S}_1} & u_{1,2}^{\mathcal{S}_1} & 1 & v_{1,1}^{\mathcal{S}_1} & v_{1,2}^{\mathcal{S}_1} & 1 \\
g_{1,1}^{\mathcal{S}_1} &g_{1,2}^{\mathcal{S}_1} &1 & 0 & 0 & 0\\
 0 & 0 & 0 & g_{2,1}^{\mathcal{S}_1} &g_{2,2}^{\mathcal{S}_1} &1
\end{bmatrix}\begin{bmatrix}
A_{(1,2,4)}^{\rm T}\\
B_{(1,2,4)}^{\rm T}
\end{bmatrix},
\end{equation}
where $A_{(1,2,4)} \triangleq [A_1,A_2,A_4] $ and other notations follow similarly. The coefficient matrix on the RHS of (\ref{eq: answer for S1}) needs to be specified according to the user demands. For $\bm{\theta}=(1,1,2)$, we let $v_{1,1}^{\mathcal{S}_1}=g_{2,1}^{\mathcal{S}_1}$ and $v_{1,2}^{\mathcal{S}_1}=g_{2,2}^{\mathcal{S}_1}$ be chosen uniformly   and independently at random from $\{0,1\}$. Also let $[u_{1,1}^{\mathcal{S}_1}; g_{1,1}^{\mathcal{S}_1}]$ and $[u_{1,2}^{\mathcal{S}_1}; g_{1,2}^{\mathcal{S}_1}]$ be chosen randomly and independently from $\{[0;1],[1;0]\}$. Note that each DB can only see the query to itself but not these to other DBs. Now we show that user 1 can decode $A_2,A_4$ and user 2 can decode $A_1,A_4$. For user 1, the interfering bits $B_1,B_2$ and $B_4$ are aligned among $A_{1,\mathcal{S}_1}^{[\bm{\theta}]}$ and $A_{2,\mathcal{S}_1,2}^{[\bm{\theta}]}$. Due the choice of the coefficients, the sub-matrix $[u_{1,2}^{\mathcal{S}_1},1; g_{1,2}^{\mathcal{S}_1},1]$ 
is full-rank (being either $[1,1;0,1]$ or  $[0,1;1,1]$ with equal probabilities). Therefore, user 1 can decode $A_2$ and $A_4$ from 
\be
\begin{bmatrix}
A_2\\
A_4
\end{bmatrix}= \begin{bmatrix}
u_{1,2}^{\mathcal{S}_1}& 1\\
g_{1,2}^{\mathcal{S}_1} &1
\end{bmatrix}^{-1} \left( \begin{bmatrix}
A_{1,\mathcal{S}_1}^{[\bm{\theta}]}-A_{2,\mathcal{S}_1,2}^{[\bm{\theta}]}\\
A_{2,\mathcal{S}_1,1}^{[\bm{\theta}]}
\end{bmatrix}   -\begin{bmatrix}
u_{1,1}^{\mathcal{S}_1}\\
g_{1,1}^{\mathcal{S}_1}
\end{bmatrix} A_1   \right)
\ee
as $A_1\in Z_1$. Similarly, user 2 can decode $A_1$ and $A_4$ since the sub-matrix $[u_{1,1}^{\mathcal{S}_1},1; g_{1,1}^{\mathcal{S}_1},1 ]$ is full-rank.

For $\mathcal{S}_2=\{1,3\}$, let $\mathcal{M}_{\{1,3\}} =\{A_1,A_3,A_5,B_1,B_3,B_5\}$. The answers are
\begin{equation}
\label{eq: answer for S2}
\begin{bmatrix}
A_{1,\mathcal{S}_2}^{[\bm{\theta}]}\\
A_{2,\mathcal{S}_2,1}^{[\bm{\theta}]}  \\
A_{2,\mathcal{S}_2,2}^{[\bm{\theta}]}  
\end{bmatrix}= 
\setlength\arraycolsep{3.7pt}
\begin{bmatrix}
u_{1,1}^{\mathcal{S}_2} & u_{1,2}^{\mathcal{S}_2} & 1 & v_{1,1}^{\mathcal{S}_2} & v_{1,2}^{\mathcal{S}_2} & 1 \\
g_{1,1}^{\mathcal{S}_2} &g_{1,2}^{\mathcal{S}_2} &1 & 0 & 0 & 0\\
 0 & 0 & 0 & g_{2,1}^{\mathcal{S}_2} &g_{2,2}^{\mathcal{S}_2} &1
\end{bmatrix}\begin{bmatrix}
A_{(1,3,5)}^{\rm T}\\
B_{(1,3,5)}^{\rm T}
\end{bmatrix}.
\end{equation} 
We let  $u_{1,1}^{\mathcal{S}_2}=g_{1,1}^{\mathcal{S}_2}$ and $v_{1,2}^{\mathcal{S}_2}=g_{2,2}^{\mathcal{S}_2}$  be chosen randomly and independently from $\{0,1\}$. Also let $[u_{1,2}^{\mathcal{S}_2}; g_{1,2}^{\mathcal{S}_2}]$ and $[v_{1,1}^{\mathcal{S}_2}; g_{2,1}^{\mathcal{S}_2}]$ be chosen randomly and independently form $\{[0;1],[1;0]\}$. From this delivery, user 1 can decode $A_3,A_5$ and user 3 can decode $B_1,B_5$. 

For $\mathcal{S}_3=\{2,3\}$, let $\mathcal{M}_{\{2,3\}} =\{A_2,A_3,A_6,B_2,B_3,B_6\}$. The answers are 
\begin{equation}
\label{eq: answer for S3}
\begin{bmatrix}
A_{1,\mathcal{S}_3}^{[\bm{\theta}]}\\
A_{2,\mathcal{S}_3,1}^{[\bm{\theta}]}  \\
A_{2,\mathcal{S}_3,2}^{[\bm{\theta}]}  
\end{bmatrix}= 
\setlength\arraycolsep{3.7pt}
\begin{bmatrix}
u_{1,1}^{\mathcal{S}_3} & u_{1,2}^{\mathcal{S}_3} & 1 & v_{1,1}^{\mathcal{S}_3} & v_{1,2}^{\mathcal{S}_3} & 1 \\
g_{1,1}^{\mathcal{S}_3} &g_{1,2}^{\mathcal{S}_3} &1 & 0 & 0 & 0\\
 0 & 0 & 0 & g_{2,1}^{\mathcal{S}_3} &g_{2,2}^{\mathcal{S}_3} &1
\end{bmatrix}\begin{bmatrix}
A_{(2,3,6)}^{\rm T}\\
B_{(2,3,6)}^{\rm T}
\end{bmatrix}.
\end{equation} 
We let  $u_{1,1}^{\mathcal{S}_3}=g_{1,1}^{\mathcal{S}_3}$ and $v_{1,2}^{\mathcal{S}_3}=g_{2,2}^{\mathcal{S}_3}$  be chosen randomly and independently from $\{0,1\}$. Also let $[u_{1,2}^{\mathcal{S}_3}; g_{1,2}^{\mathcal{S}_3}]$ and $[v_{1,1}^{\mathcal{S}_3}; g_{2,1}^{\mathcal{S}_3}]$ be chosen randomly and independently form $\{[0;1],[1;0]\}$.
From this delivery, user 2 can decode $A_3,A_6$ and user 3 can decode $B_2,B_6$. 

\textit{\underline{Decodability}:} From each $\mathcal{S}_i,i\in[3]$, the users therein can directly decode some desired bits. More specifically, from $\mathcal{S}_1$ and $\mathcal{S}_2$, user 1 decodes $A_2,A_3,A_4,A_5$ and still needs $A_6$ which can be decoded from $A_{2,\mathcal{S}_3,1}^{[\bm{\theta}]}$ by subtracting the already obtained bits $A_2$ and $A_3$, i.e.,  $A_6= A_{2,\mathcal{S}_3,1}^{[\bm{\theta}]}-g_{1,1}^{\mathcal{S}_3}A_2 -g_{1,2}^{\mathcal{S}_3}A_3$. The decodability of $A_6$ is guaranteed because  in (\ref{eq: answer for S3}), the coefficient of $A_6$ is always equal to $1$ regardless of $\bm{\theta}$. From $\mathcal{S}_1$ and $\mathcal{S}_3$, user 2 decodes $A_1,A_4,A_3,A_6$ and further decodes $A_5$ from $A_{2,\mathcal{S}_2,1}^{[\bm{\theta}]}$. Similarly, user 3 decodes $B_1,B_2,B_5,B_6$ from $\mathcal{S}_2$ and $\mathcal{S}_3$ and  further decodes $B_4$ from $  A_{2,\mathcal{S}_1,2}^{[\bm{\theta}]} $. Therefore, all users can recover their desired messages. 

\textit{\underline{Privacy}:} First, the delivery for each user subset $\mathcal{S}$ is private, i.e., each DB can not tell which bit in $\mathcal{M}_{\mathcal{S}}$ is needed by which user in $\mathcal{S}$. Second, the delivery for different subsets are independent. These two aspects lead to the privacy of the overall delivery scheme. We now give a brief explanation of why the delivery for each subset is private.  Taking $\mathcal{S}_1$ for instance, demand  privacy requires that each DB can not determine if the user demands $(\theta_1,\theta_2)$ are $(A_2,A_1),(A_2,B_1),(B_2,A_1)$ or $(B_2,B_1)$. For example, from DB 1's perspective, the linear combination coefficients $u_{1,1}^{\mathcal{S}_1},u_{1,2}^{\mathcal{S}_1},v_{1,1}^{\mathcal{S}_1}$ and $v_{1,2}^{\mathcal{S}_1}$ in $A_{1,\mathcal{S}_1}^{[\bm{\theta}]}$ are chosen randomly and independently as 0 or 1. As DB 1 does not know which bits (in $A_1,A_2,B_1,B_2$) are aligned among $A_{1,\mathcal{S}_1}^{[\bm{\theta}]}$ and  $A_{2,\mathcal{S}_1,1}^{[\bm{\theta}]},A_{2,\mathcal{S}_1,2}^{[\bm{\theta}]}$, it can not obtain any information about which bits are decoded by user 1 and 2 respectively. Hence, the delivery is private from DB 1's viewpoint. Similarly, the delivery is also private from DB 2's viewpoint. 

\textit{\underline{Performance}:} As $D=9$ bits are downloaded in total, the achieved load is $R=3/2$. 
\hfill $\lozenge$
\end{example}

\begin{example} 
\label{example (1,3/4)}
(\textbf{Achievability of $(1,3/4)$}) Consider the same problem setting as Example \ref{example (1/3,3/2)}. We consider the memory-load pair corresponding to $t=2$ in~\eqref{eq: thm 1 pairs}, which is $(1,3/4)$.

\textit{1) Cache placement:}  Assume that each message consists of $L=4$ bits, i.e., $A=(A_1,A_2,A_3,A_4)$, $B=(B_1,B_2,B_3,B_4)$. Each user stores two bits from each message, i.e., $Z_1=\{A_2,A_3,B_2,B_3\}, Z_2=\{A_1,A_3,B_1,B_3\}$ and $Z_3=\{A_1,A_2,B_1,B_2\}$. Therefore,  $M=1$.

\textit{2) Private delivery:} Suppose $\bm{\theta}=(1,2,2)$. The answer of DB 1 contains one linear combination and the answer of DB 2 contains two linear combinations, i.e., $A_{2}^{[\bm{\theta}]}=(A_{2,1}^{[\bm{\theta}]}, A_{2,2}^{[\bm{\theta}]})$. Since $t=2$, there is only one user subset of size $t+1=3$ which is $\{1,2,3\}$, we ignore the superscript in the linear coefficients. The answers are
\be
\begin{bmatrix}
A_{1}^{[\bm{\theta}]}\\
A_{2,1}^{[\bm{\theta}]}\\
A_{2,2}^{[\bm{\theta}]}
\end{bmatrix} =
\setlength\arraycolsep{3pt}
\begin{bmatrix}
u_{1,1} & u_{1,2}& u_{1,3} & 1 & v_{1,1} &v_{1,2} & v_{1,3}& 1 \\
g_{1,1} & g_{1,2} & g_{1,3} &1 &0&0&0&0\\
0&0&0&0&    g_{2,1}& g_{2,2} & g_{2,3} & 1
\end{bmatrix}
\begin{bmatrix}
A_{(1:4)}^{\rm T}\\
B_{(1:4)}^{\rm T}
\end{bmatrix}.
\ee
For $\bm{\theta}=(1,2,2)$, we let $u_{1,2}=g_{1,2}, u_{1,3}=g_{1,3}$ and $v_{1,1}=g_{2,1}$ be chosen uniformly   and independently at random from $\{0,1\}$. Also let $[u_{1,1};g_{1,1}], [v_{1,2};g_{2,2}]$ and $[v_{1,3};g_{2,3}]$ be chosen randomly and independently from $\{[0;1],[1;0]\}$. 

\textit{\underline{Decodability}:} For user 1, it needs $A_1$ and $A_4$ since the other two bits have been cached. As $B_1$ and $B_4$ are aligned among $A_1^{[\bm{\theta}]}$ and $A_{2,2}^{[\bm{\theta}]}$, we have $v_{1,1}B_1+B_4=g_{2,1}B_1+B_4 = A_{2,2}^{[\bm{\theta}]}-g_{2,2}B_2-g_{2,3}B_3$. Due to the choice of $[u_{1,1};g_{1,1}]$, the sub-matrix $[u_{1,1},1; g_{1,1},1]$ is full-rank. Hence,  user 1 can decode $A_1$  and $A_4$ from 
\be
\begin{bmatrix}
A_1\\
A_4
\end{bmatrix}  =
\begin{bmatrix}
u_{1,1} & 1\\
g_{1,1} & 1 
\end{bmatrix}^{-1}\left( \mathbf{X}\begin{bmatrix}
A_2\\A_3
\end{bmatrix} -\mathbf{X}'\begin{bmatrix}
B_2\\
B_3
\end{bmatrix} - \yv  \right),
\ee
where $\mathbf{X}=[u_{1,2},u_{1,3};g_{1,2},g_{1,3}]$, $\mathbf{X}'=[v_{1,2},v_{1,3};0,0]$ and  $\yv =[A_{2,2}^{[\bm{\theta}]}-g_{2,2}B_2-g_{2,3}B_3;0]$.
Similarly, due to the alignment of $A_2,A_4$ among $A_1^{[\bm{\theta}]}$ and $A_{2,1}^{[\bm{\theta}]}$, user 2 can decode $B_2$ and $B_4$. Also, due to the alignment of $A_3,A_4$ among $A_1^{[\bm{\theta}]}$ and $A_{2,1}^{[\bm{\theta}]}$, user 3 can decode $B_3$ and $B_4$. Therefore, all users can recover their desired messages. 

\textit{\underline{Privacy}:} From DB 1's viewpoint, the linear coefficients $u_{1,1},u_{1,2},u_{1,3}$, $v_{1,1},v_{1,2}$ and $v_{1,3}$ in $A_1^{[\bm{\theta}]}$ are chosen randomly and independently as 0 or 1. As DB 1 does not know which bits are aligned among $A_1^{[\bm{\theta}]}$ and $A_{2,1}^{[\bm{\theta}]},A_{2,2}^{[\bm{\theta}]}$, it can not determine which bits are decoded by which user. Therefore, the delivery is private from DB 1's perspective. Similarly, the scheme is also private from DB 2's perspective.

\textit{\underline{Performance}:} As $D=3$ bits are downloaded, the achieved load is $R=3/4$.
\hfill $\lozenge$
\end{example}

\section{ General Achievable Schemes}
\label{section: proof of thm 1}
In this section, we present the general achievable schemes for the MuPIR problem with  $K=2$ messages, $K_{\rm u}\ge 2$ users and $N\ge 2$ databases to prove Theorem~\ref{thm 1}.

Let $t\in [K_{\rm u}-1]$. Assume that each message consists of  $L=\binom{K_{\rm u}}{t}(N-1)+\binom{K_{\rm u}}{t+1}$ bits.\footnote{It can be seen that $L$ is linear in $N$ while in the product design, a quadratic message size $\binom{K_{\rm u}}{t}N^2$ is required. This is another advantage of the proposed scheme compared to the product design.} We split each message into $\binom{K_{\rm u}}{t}$ \textit{blocks} plus $\binom{K_{\rm u}}{t+1}$ \emph{extra bits}. Each message block contains $N-1$ bits. Therefore, each message $W_k$ can be written as 
\begin{align}
W_k = &\left\{ W_{k,\mathcal{T}}^{[1:N-1]}: \forall \mathcal{T}\in \binom{[K_{\rm u}]}{t} \right  \}\nonumber\\
  &  \quad \;\cup \left\{W_k^{\mathcal{S}}:\forall \mathcal{S}\in \binom{[K_{\rm u}]}{t+1}  \right\},\; k=1,2,
\end{align}
in which each $W_{k,\mathcal{T}}^{[1:N-1]} \triangleq \{W_{k,\mathcal{T}}^b:\forall b\in[N-1]\}$ represents a message block and each $W_{k}^{\mathcal{S}}$ represents an extra bit. With this message splitting, the  cache placement and private delivery phases are described as follows. 

\textit{1) Cache placement:} Each user $u$ stores $\binom{K_{\rm u}-1}{t-1}$ blocks of each message, i.e., the cache of user $u$ is 
\be
Z_u =\left\{ W_{k,\mathcal{T}}^{[1:N-1]}: \forall \mathcal{T}\in \binom{[K_{\rm u}]}{t}, u\in\mathcal{T},\forall k\in[2] \right\}. 
\ee
Therefore, $M= 2\binom{K_{\rm u}-1}{t-1}(N-1) /L$.

\textit{2) Private delivery:} 
The private delivery phase consists of $\binom{K_{\rm u}}{t+1}$ steps each corresponding to a   subset  of  $t+1$ users (denoted by $\mathcal{S}$) and a   set of message bits $\mathcal{M}_{\mathcal{S}}$. 
The answer of DB $n\in[N]$ consists of $\binom{K_{\rm u}}{t+1}$ parts, i.e., $A_n^{[\bm{\theta}]}=\left(A_{n,\mathcal{S}}^{[\bm{\theta}]}:\forall \mathcal{S} \in \binom{[K_{\rm u}]}{t+1}    \right)$. Each $A_{n,\mathcal{S}}^{[\bm{\theta}]},n\in[N-1]$ is a linear combination of the message bits in $\mathcal{M}_{\mathcal{S}}$ and each $A_{N,\mathcal{S}}^{[\bm{\theta}]}$ contains two linear combinations of $\mathcal{M}_{\mathcal{S}}$. More precisely, let  
$$
\mathcal{M}_{\mathcal{S}} =\left\{ W_{k,\mathcal{S}\backslash \{u\}}^{[1:N-1]}: \forall u\in \mathcal{S}, \forall k\in[2]    \right\}\cup \left\{  W_k^{\mathcal{S}}: \forall k\in[2]  \right\},
$$
i.e., $\mathcal{M}_{\mathcal{S}}$ contains $t+1$ message blocks and one extra bit of each message. 
All the bits in $\left\{  W_{k,\mathcal{S}\backslash\{u'\}}^{[1:N-1]}: u\ne u',\forall k\in[2] \right\}$ have been cached by user $u\in \mathcal{S}$, and thus it needs to recover the block $W_{\theta_u,\mathcal{S}\backslash\{u\}}^{[1:N-1]}$ and the extra bit $W_{\theta_u}^{\mathcal{S}}$ in the delivery phase.
The answers corresponding to each $\mathcal{S}$ is shown in~\eqref{answer: general scheme} at the top of the next page. Suppose $\mathcal{S}=\{u_1,u_2,\ldots,u_{t+1}\}$, the message bits in $\mathcal{M}_{\mathcal{S}}$ on the RHS of (\ref{answer: general scheme}) are arranged in the following way $\forall k\in[2]$:
$$
\left( W_{k,\mathcal{S}\backslash \{u\}}^{(1:N-1)}:\forall u\in\mathcal{S}    \right)^{\rm T}\triangleq \left[
W_{k,\mathcal{S}\backslash \{u_1\}}^{(1:N-1)},\ldots, W_{k,\mathcal{S}\backslash \{u_{t+1}\}}^{(1:N-1)} 
 \right]^{\rm T},
$$
where $ W_{k,\mathcal{S}\backslash \{u\}}^{(1:N-1)} \triangleq [ W_{k,\mathcal{S}\backslash \{u\}}^{1},\ldots,W_{k,\mathcal{S}\backslash \{u\}}^{N-1}]$ is an ordered representation of the message block $W_{k,\mathcal{S}\backslash \{u\}}^{[1:N-1]}$.
\begin{figure*}[t]
\be 
\label{answer: general scheme}
\begin{bmatrix}
A_{1,\mathcal{S}}^{[\bm{\theta}]}\\
A_{2,\mathcal{S}}^{[\bm{\theta}]}\\
\vdots \\
A_{N-1,\mathcal{S}}^{[\bm{\theta}]}\\
A_{N,\mathcal{S},1}^{[\bm{\theta}]}\\
A_{N,\mathcal{S},2}^{[\bm{\theta}]}
\end{bmatrix} =
\setlength\arraycolsep{3.6pt}
\underbrace{\begin{bmatrix}
\uv_{1,1}^{\mathcal{S}}&\uv_{1,2}^{\mathcal{S}} &\cdots & \uv_{1,t+1}^{\mathcal{S}} & 1& \vv_{1,1}^{\mathcal{S}}&\vv_{1,2}^{\mathcal{S}} &\cdots & \vv_{1,t+1}^{\mathcal{S}} & 1\\
\uv_{2,1}^{\mathcal{S}}&\uv_{2,2}^{\mathcal{S}} &\cdots & \uv_{2,t+1}^{\mathcal{S}} & 1& \vv_{2,1}^{\mathcal{S}}&\vv_{2,2}^{\mathcal{S}} &\cdots & \vv_{2,t+1}^{\mathcal{S}} & 1\\
\vdots &\vdots & \ddots& \vdots &\vdots &\vdots & \vdots & \ddots & \vdots & \vdots\\
\uv_{N-1,1}^{\mathcal{S}}&\uv_{N-1,2}^{\mathcal{S}} &\cdots & \uv_{N-1,t+1}^{\mathcal{S}} & 1& \vv_{N-1,1}^{\mathcal{S}}&\vv_{N-1,2}^{\mathcal{S}} &\cdots & \vv_{N-1,t+1}^{\mathcal{S}} & 1\\
\bm{g}_{1,1}^{\mathcal{S}}&\bm{g}_{1,2}^{\mathcal{S}} &\cdots & \bm{g}_{1,t+1}^{\mathcal{S}} & 1& \bm{0}_{N-1}&\bm{0}_{N-1} &\cdots &\bm{0}_{N-1}& 0 \\
\bm{0}_{N-1}&\bm{0}_{N-1} &\cdots &\bm{0}_{N-1}& 0 & \bm{g}_{2,1}^{\mathcal{S}}&\bm{g}_{2,2}^{\mathcal{S}} &\cdots & \bm{g}_{2,t+1}^{\mathcal{S}} & 1
\end{bmatrix}}_{\textrm{dimension: } (N+1)\times 2\left(   (t+1)(N-1)+1\right)}
\begin{bmatrix}
\left( W_{1,\mathcal{S}\backslash \{u\}}^{(1:N-1)}:\forall u\in\mathcal{S}    \right)^{\rm T} \\
W_1^{\mathcal{S}}\\
\left( W_{2,\mathcal{S}\backslash \{u\}}^{(1:N-1)}:\forall u\in\mathcal{S}    \right)^{\rm T} \\
W_2^{\mathcal{S}}
\end{bmatrix}
\ee
\end{figure*}

We next show how to design the  linear coefficients $\uv_{n,j}^{\mathcal{S}},\vv_{n,j}^{\mathcal{S}},\bm{g}_{k,j}^{\mathcal{S}}\in \mathbb{F}_2^{1\times (N-1)} ,\forall n,j,k$ according to $\bm{\theta}$.  
For any $u\in[K_{\rm u}]$, let $\bar{\theta}_u $ be the complement of $\theta_u$, i.e., $\{\bar{\theta}_u,\theta_u\}=\{1,2\}$. For each $i\in[t+1]$, define two $N\times N$ coefficient sub-matrices of~\eqref{answer: general scheme} as
\be 
\mathbf{U}_i^{\mathcal{S}}  \triangleq \begin{bmatrix}
\uv_{1,i}^{\mathcal{S}}& 1\\
\vdots &\vdots \\
\uv_{N-1,i}^{\mathcal{S}}& 1\\
\bm{g}_{1,i}^{\mathcal{S}}& 1 
\end{bmatrix},\quad 
\mathbf{V}_i^{\mathcal{S}}  \triangleq 
\begin{bmatrix}
\vv_{1,i}^{\mathcal{S}}& 1\\
\vdots& \vdots\\
\vv_{N-1,i}^{\mathcal{S}}& 1\\
\bm{g}_{2,i}^{\mathcal{S}}& 1 
\end{bmatrix}.
\ee
It can be seen that $\mathbf{U}_i^{\mathcal{S}}$ and $\mathbf{V}_{i}^{\mathcal{S}}$ correspond to the $i$-th message block and the extra bit $W_k^{\mathcal{S}}$ of the two messages in $\mathcal{M}_{\mathcal{S}}$, respectively. Moreover, denote 
$
\mathbf{Y}_n \triangleq \setlength\arraycolsep{2pt} \begin{bmatrix}
\mathbf{I}_{n-1}& \bm{1}_{n-1}^{\rm T}\\
\bm{0}_{n-1} & 1
\end{bmatrix} 
$
where $n\geq 2$
and let $\mathcal{Y}_{n}$ be a set containing the $n$ rows of $\mathbf{Y}_n$. It can be seen that $\mathbf{Y}_n$ has full rank for any $n$.

Suppose $\mathcal{S}=\{u_1,u_2,\ldots,u_{t+1}\}$. Define two subsets $\mathcal{I}_1,\mathcal{I}_2$ of $[t+1]$ such that $\mathcal{I}_1 \cup \mathcal{I}_2=[t+1]$ and $\mathcal{I}_1\cap \mathcal{I}_2=\emptyset$. 
Suppose that $\theta_{u_i}=1,\forall i\in\mathcal{I}_1$ and $\theta_{u_i}=2,\forall i\in\mathcal{I}_2$. Let $\mathcal{S}_{\mathcal{I}_1}=\{u_i:\forall i\in\mathcal{I}_1\}$ and $\mathcal{S}_{\mathcal{I}_2}=\{u_i:\forall i\in\mathcal{I}_2\}$ be the two subsets of $\mathcal{S}$ in which the users demand $W_1$ and $W_2$ respectively ($\mathcal{S}_{\mathcal{I}_1} \cup \mathcal{S}_{\mathcal{I}_2}=\mathcal{S}$). The linear coefficients are specified as follows.

{The general idea is that for each user in $\mathcal{S}$, we align the interfering message bits (which are neither requested nor cached by this user) among the  $N+1$ linear combinations in~\eqref{answer: general scheme},  such that they can be removed and the users can decode the $N$ desired bits from the remaining $N$ linear combinations.} In particular, for each $i\in \mathcal{I}_1$, we let 
$  \vv_{1,i}^{\mathcal{S}}=\cdots=\vv_{N-1,i}^{\mathcal{S}}=\bm{g}_{2,i}^{\mathcal{S}}
$ be chosen uniformly at random  from $\mathcal{Y}_N$ and let $\mathbf{U}_i^{\mathcal{S}}$ be a random permutation of the rows of $\mathbf{Y}_N$.
For each $i'\in \mathcal{I}_2$, we let 
$  \uv_{1,i'}^{\mathcal{S}}=\cdots=\uv_{N-1,i'}^{\mathcal{S}}=\bm{g}_{1,i'}^{\mathcal{S}}
$ be chosen uniformly at random from $\mathcal{Y}_N$ and let $\mathbf{V}_{i'}^{\mathcal{S}}$ be a random permutation of the rows of $\mathbf{Y}_N$.
The choices for all $i\in \mathcal{I}_1,i'\in\mathcal{I}_2$ are independent.

\textit{\underline{Decodability}:} For any user $u\in[K_{\rm u}]$, in the delivery phase it needs to recover totally $\binom{K_{\rm u}-1}{t}$ message blocks (i.e.,  $ \{W_{\theta_u,\mathcal{T}}^{[1:N-1]}: u\notin \mathcal{T} \}$), and all the $\binom{K_{\rm u}}{t+1}$ extra bits (i.e., $\{W_{\theta_u}^{\mathcal{S}}:\forall \mathcal{S} \}$). We now show that with the above delivery design, user $u$ can correctly decode all the desired bits of $W_{\theta_u}$ in two steps. 

In the first step, from each $\mathcal{S}\in \binom{[K_{\rm u}]}{t+1}$ for which $u\in \mathcal{S}$, user $u$ can decode the block $W_{\theta_u,\mathcal{S}\backslash\{u\}}^{[1:N-1]}$ and the extra bit $W_{\theta_u}^{\mathcal{S}}$. The reason is as follows. Note that the blocks $\{W_{\theta_u,\mathcal{S}\backslash \{u'\}}^{[1:N-1]}: \forall u'\in \mathcal{S}\backslash \{u\} \}$ have already been cached by user $u$ and can be removed from the linear combinations in~\eqref{answer: general scheme}. 
Now the RHS of~\eqref{answer: general scheme} only involves two message blocks $W_{\theta_u,\mathcal{S}\backslash \{u\}}^{[1:N-1]}, W_{\bar{\theta}_u,\mathcal{S}\backslash \{u\}}^{[1:N-1]}$ and two extra bits $W_{\theta_u}^{\mathcal{S}},W_{\bar{\theta}_u}^{\mathcal{S}}$. {From the  design of the coefficients,  $W_{\bar{\theta}_u,\mathcal{S}\backslash \{u\}}^{[1:N-1]}$ and $W_{\bar{\theta}_u}^{\mathcal{S}}$ are aligned among $A_{n,\mathcal{S}}^{[\bm{\theta}]},\forall n\in[N]$. In addition, using the answer from DB $N$, $W_{\bar{\theta}_u,\mathcal{S}\backslash \{u\}}^{[1:N-1]}$ and $W_{\bar{\theta}_u}^{\mathcal{S}}$ can be subtracted from $A_{n,\mathcal{S}}^{[\bm{\theta}]},\forall n\in[N-1]$. Then user $u$
can recover the $N$ desired  bits (i.e., $W_{\theta_u,\mathcal{S}\backslash\{u\}}^{[1:N-1]},W_{\theta_u}^{\mathcal{S}}$) from the remaining $N$ linear combinations in~\eqref{answer: general scheme} . This is because  in these $N $ linear combinations, the $N\times N$ coefficient sub-matrix corresponding to the $N$ desired bits is a random permutation of the rows of $\mathbf{Y}_N$,  and thus  is full-rank.} By considering all $\mathcal{S} \subseteq [K_{\rm u}]$ where $|\mathcal{S}|=t+1$ and $u\in\mathcal{S}$, user $u$ can decode the $\binom{K_{\rm u}-1}{t}$ desired message blocks and $\binom{K_{\rm u}-1}{t}$ desired extra bits.

In the second step, user $u$ decodes an extra bit $W_{\theta_u}^{\mathcal{S}}$ from {each $\mathcal{S} \subseteq [K_{\rm u}]$ where $|\mathcal{S}|=t+1$ and $u\notin\mathcal{S}$}, using the decoded message blocks in the first step. More specifically, for each $\mathcal{S}=\{u_1,\ldots,u_{t+1}\}\in\binom{[K_{\rm u}]\backslash\{u\}}{t+1}$, the answer $A_{N,\mathcal{S}}^{[\bm{\theta}]}$ of DB $N$ contains two linear combinations $ \forall k\in[2]:$
\be 
\label{eq: DB N linear combination}
A_{N,\mathcal{S},k}^{[\bm{\theta}]} =  \sum_{i=1}^{t+1} \bm{g}_{k,i} \left[ W_{k,\mathcal{S}\backslash\{u_i\}}^1,\ldots, W_{k,\mathcal{S}\backslash\{u_i\}}^{N-1} \right]^{\rm T}+ W_{k}^{\mathcal{S}}.
\ee
As the blocks $W_{\theta_u, \mathcal{S}\backslash\{u_i\}}^{[1:N-1]},\forall i\in[t+1]$ have already been decoded by user $u$ in the first step, the extra bit $W_{\theta_u}^{\mathcal{S}}$ can be decoded from (\ref{eq: DB N linear combination}) by subtracting $\sum_{i=1}^{t+1}\bm{g}_{k,i} [ W_{k,\mathcal{S}\backslash\{u_i\}}^1,\ldots, W_{k,\mathcal{S}\backslash\{u_i\}}^{N-1}]^{\rm T}$ from $A_{N,\mathcal{S},\theta_u}^{[\bm{\theta}]}$. 
As a result, user $u$ decodes all the desired bits of $W_{\theta_u}$. 

\textit{\underline{Privacy}:} First, the delivery for each subset $\mathcal{S}$ is private. Second, because the specifications of the linear coefficients for different subsets are independent, the delivery scheme is private. Next we show why the delivery for each subset is private. From the viewpoint of DB $n\in[N-1]$, the linear coefficients $\uv_{n,1}^{\mathcal{S}},\ldots,\uv_{n,t+1}^{\mathcal{S}},   \vv_{n,1}^{\mathcal{S}},\ldots,\vv_{n,t+1}^{\mathcal{S}}$ are chosen randomly and independently from $\mathcal{Y}_N$ (excluding the last entry 1 of each element in $\mathcal{Y}_N$). The linear coefficients $\bm{g}_{k,i}^{\mathcal{S}},\forall i\in[t+1],\forall k\in[2]$ also appear to be randomly and independently distributed over $\mathcal{Y}_N$ from DB $N$'s viewpoint. As each DB $n\in[N]$ does not know which message blocks are aligned among the $N$ answers, it can not determine which message block is decoded by which user. Therefore, the delivery is private. 

\textit{\underline{Performance}:} As the answers corresponding to each $\mathcal{S}$ contains $N+1$ bits, $D=\binom{K_{\rm u}}{t+1}(N+1)$ bits are downloaded in total, achieving the load  $R={\binom{K_{\rm u}}{t+1}(N+1)}/{L }$. 

\section{Conclusion}
We studied the cache-aided MuPIR problem and proposed a new scheme for two messages, arbitrary number of users and databases with uncoded prefetching. The proposed scheme uses a novel message block based cache placement method and a user group based delivery scheme. By exploiting the linear combinations among the answers for multiple user subsets, the proposed scheme achieves better performance than any previously proposed schemes including the product design.

\appendices
\section{Proof of Corollary \ref{corollary: comparison to PD}}
\label{appendix: proof Corollary 1}
Let $R^{\rm PD}$ denote the load achieved by the product design, i.e., the lower convex envelope of the memory-load pairs $(0,2)$  and $(M^{\rm PD}_t,R^{\rm PD}_t),\forall t\in [K_{\rm u}]$ where $M^{\rm PD}_t= {tK}/{K_{\rm u}}$ and 
\be
\label{eq: PD load}
R^{\rm PD}_t = \min\left\{\frac{K_{\rm u}-t}{t+1}\left( 1+\frac{1}{N} \right),K-M_t^{\rm PD}     \right\}.
\ee 
We next prove that the load $R$ in Theorem \ref{thm 1} is smaller than $R^{\rm PD}$ for all $M\in[0,2]$. Since both $R$ and $R^{\rm PD}$ take value $K-M$ when the naive scheme achieves a lower load, we only need to compare the load $R$ implied by the lower convex envelope of $(M_t,R_t),\forall t\in[K_{\rm u}-1]$ where $R_t=\frac{\binom{K_{\rm u}}{t+1}(N+1)}{\binom{K_{\rm u}}{t}(N-1)+\binom{K_{\rm u}}{t+1}}$ with the load $R^{\rm PD}$ implied by the lower convex envelope of $(M^{\rm PD}_t,R^{\rm PD}_t),\forall t\in[K_{\rm u}]$ where $R^{\rm PD}=\frac{K_{\rm u}-t}{t+2}(1+\frac{1}{N})$ ($M^{\rm PD}_t=tK/K_{\rm u}$ denotes the memory size corresponding to $t$ in the product design). It can be easily seen that both the load curves $R$ and $R^{\rm PD}$ are convex when $t\ge 1$ (not considering the naive scheme load term). For clarity, we use $R(M)$ and $R^{\rm PD}(M)$ to denote the load at memory size $M$ for both schemes in the following. It can be seen that $R_t=R(M_t),R^{\rm PD}_t=R(M^{\rm PD}_t),\forall t$.
Note that $M_t\le M^{\rm PD}_t,\forall t\in[K_{\rm u}-2]$ which is due to 
\be
M_t=\frac{K\binom{K_{\rm u}-1}{t-1}(N-1)   }{\binom{K_{\rm u}}{t}(N-1)+\binom{K_{\rm u}}{t+1}}\le \frac{K\binom{K_{\rm u}-1}{t-1}(N-1)   }{\binom{K_{\rm u}}{t}(N-1)} =M^{\rm PD}_t.
\ee

We next prove that $R(M)\le R^{\rm PD}(M),\forall M_t\le M\le M_{t+1}$ and $\forall t\in[K_{\rm u}-2]$. Note that when $M\ge M_{K_{\rm u}-1}$, the proposed scheme is optimal by Corollary \ref{corollary: optimality at large memory regime}, so $R(M)\le R^{\rm PD}(M)$ when $M\ge M_{K_{\rm u}-1}$. Because $R^{\rm PD}(M)$ is convex, proving that $R(M)\le \widehat{R}_t^{\rm PD}(M)$ and $R(M)\le \widehat{R}_t^{\rm PD}(M),\forall M_t\le M\le M_{t+1},\forall t$ is sufficient to guarantee $R(M)\le R^{\rm PD}(M),\forall 0\le M\le K$. $\widehat{R}_t^{\rm PD}(M)$ represents the line crossing the two product design points $\left(M^{\rm PD}_t, R^{\rm PD}(M^{\rm PD}_t)\right)$ and $\left(M^{\rm PD}_{t+1}, R^{\rm PD}(M^{\rm PD}_{t+1})\right)$, i.e., $\forall 0\le M\le 2$,
\be
 \widehat{R}_t^{\rm PD}(M) = \frac{1+N^{-1}}{t+1} \left(2K_{\rm u}-t+1 -\frac{K_{\rm u}(K_{\rm u}+1)}{tK} M        \right).
\ee
Since $R(M)$ is linear when $M_{t}\le M\le M_{t+1}$, proving  $R(M)\le \widehat{R}_t^{\rm PD}(M)$ and $R(M)\le \widehat{R}_t^{\rm PD}(M),\forall M_t\le M\le M_{t+1}$ is equivalent to proving that $R(M_t)\le \widehat{R}_t^{\rm PD}(M_t)$ and $R(M_{t+1})\le \widehat{R}_t^{\rm PD}(M_{t+1}),\forall t\in[K_{\rm u}-1]$ which is shown as follows.

We first prove $R(M_t)\le \widehat{R}_t^{\rm PD}(M_t),\forall t\in[K_{\rm u}-2]$. Denote $\forall t\in[K_{\rm u}-2]$:
\be
\alpha_t \triangleq \frac{\binom{K_{\rm u}}{t+1}}{\binom{K_{\rm u}}{t}(N-1)},\quad \beta_t\triangleq \frac{\binom{K_{\rm u}}{t+1}-\binom{K_{\rm u}}{t}   }{\binom{K_{\rm u}}{t}N}.
\ee
Plugging in $M_t =\frac{tK}{K_{\rm u}(1+\alpha_t)}$, we have 
\begin{align}
&\widehat{R}_t^{\rm PD}(M_t)=\frac{1+N^{-1}}{t+1}\left( 2K_{\rm u}-t+1 -\frac{K_{\rm u}+1}{1+\alpha_t} \right),\\
&R(M_t) = \frac{(1+N^{-1})(K_{\rm u}-t)}{(1+\beta_t)(t+1)}.
\end{align}
Let $\gamma_{t,N}(M_t)\triangleq \frac{\widehat{R}_t^{\rm PD}(M_t)}{R(M_t)}$ denote the ratio of the loads between PD and the proposed scheme at $M=M_t$. Then
\begin{subequations}
\begin{align}
&\gamma_{t,N}(M_t)\\
&= \frac{1+\beta_t}{K_{\rm u}-t}\left( 2K_{\rm u}-t+1 -\frac{K_{\rm u}+1}{1+\alpha_t} \right)\\
&=\frac{2K_{\rm u}-t+1 }{K_{\rm u}-t}(1+\beta_t) -\frac{(K_{\rm u}+1)(1+\beta_t)}{(K_{\rm u}-1)(1+\alpha_t)}\\
&=\left(1+\frac{K_{\rm u}+1}{K_{\rm u}-t}\right)\left(1+ \frac{1}{N}\left( \frac{K_{\rm u}+1}{t+1} -2\right)   \right)\nonumber\\
&\quad + \frac{K_{\rm u}+1}{K_{\rm u}-t}\frac{1+\left(   \frac{K_{\rm u}+1}{t+1}-2\right) N^{-1}  }{1+\left(   \frac{K_{\rm u}+1}{t+1}-1\right)(N-1)^{-1}  }\\
&=\left(  \left( 1+ \frac{K_{\rm u}+1}{K_{\rm u}-t} \right)\left(  \frac{K_{\rm u}+1}{t+1}-2 \right) +\frac{K_{\rm u}+1}{K_{\rm u}-t}     \right)\frac{1}{N}+1\\
&= \left( \frac{K_{\rm u}+1}{t+1} \left(1+\frac{K_{\rm u}+1}{K_{\rm u}-t}  \right) - \frac{K_{\rm u}+1}{K_{\rm u}-t}-2 \right)\frac{1}{N} + 1\\
& =2\left(\frac{K_{\rm u}+1}{t+1}-1  \right)\frac{1}{N}+1.
\end{align}
\end{subequations}
Since $t\le K_{\rm u}-2$, we have $\frac{K_{\rm u}+1}{t+1}-1\ge \frac{K_{\rm u}+1}{K_{\rm u}-1}-1=\frac{2}{K_{\rm u}-1}>0$, implying $\gamma_{t,N}(M_t)\ge 1$, i.e., $\widehat{R}_t^{\rm PD}(M_t)\ge R(M_t),\forall t\in[K_{\rm u}-2]$. Similarly, by replacing $t$ with $t+1$, we have $\gamma_{t+1,N}(M_{t+1})\triangleq \frac{\widehat{R}_{t+1}^{\rm PD}(M_{t+1})}{R(M_{t+1})}\ge 1$ where $M_{t+1}=\frac{(t+1)K}{K_{\rm u}(1+\alpha_{t+1})}$ and $\widehat{R}_{t+1}^{\rm PD}(M)$ represents the line crossing the two points $\left(M_{t+1}^{\rm PD} ,  R^{\rm PD}(M_{t+1}^{\rm PD})\right)$ and $\left(M_{t+2}^{\rm PD} ,  R^{\rm PD}(M_{t+2}^{\rm PD})\right),\forall t\in[K_{\rm u}-3]$. Recall that $M_t\le M^{\rm PD}_t,\forall t\in[K_{\rm u}-2]$. Because $R^{\rm PD}(M)$ is convex on $M_1\le M \le K$,  we have $R^{\rm PD}_{t+1}(M)\le R^{\rm PD}_t(M), \forall M\le M_{t+1}$, which implies that $R(M_{t+1})\le  R^{\rm PD}_t(M_{t+1})$. As a result, we proved that $R(M)\le R^{\rm PD}(M),\forall 0\le M\le K(K=2)$.

It was proved by \cite{9174083} that the product design load $R^{\rm PD}$ of \eqref{eq: PD load} is optimal within a factor 8. Since the load in Theorem \ref{thm 1} is smaller than $R^{\rm PD}$, it is also optimal within a factor of 8. Therefore, the proof of Corollary  \ref{corollary: comparison to PD} is complete.

\bibliographystyle{IEEEtran}
\bibliography{references_d2d}

\begin{thebibliography}{10}
\providecommand{\url}[1]{#1}
\csname url@samestyle\endcsname
\providecommand{\newblock}{\relax}
\providecommand{\bibinfo}[2]{#2}
\providecommand{\BIBentrySTDinterwordspacing}{\spaceskip=0pt\relax}
\providecommand{\BIBentryALTinterwordstretchfactor}{4}
\providecommand{\BIBentryALTinterwordspacing}{\spaceskip=\fontdimen2\font plus
\BIBentryALTinterwordstretchfactor\fontdimen3\font minus
  \fontdimen4\font\relax}
\providecommand{\BIBforeignlanguage}[2]{{%
\expandafter\ifx\csname l@#1\endcsname\relax
\typeout{** WARNING: IEEEtran.bst: No hyphenation pattern has been}%
\typeout{** loaded for the language `#1'. Using the pattern for}%
\typeout{** the default language instead.}%
\else
\language=\csname l@#1\endcsname
\fi
#2}}
\providecommand{\BIBdecl}{\relax}
\BIBdecl

\bibitem{chor1995private}
B.~Chor, O.~Goldreich, E.~Kushilevitz, and M.~Sudan, ``Private information
  retrieval,'' in \emph{Proceedings of IEEE 36th Annual Foundations of Computer
  Science}.\hskip 1em plus 0.5em minus 0.4em\relax IEEE, 1995, pp. 41--50.

\bibitem{sun2017capacity}
H.~Sun and S.~A. Jafar, ``The capacity of private information retrieval,''
  \emph{IEEE Transactions on Information Theory}, vol.~63, no.~7, pp.
  4075--4088, 2017.

\bibitem{tandon2017capacity}
R.~Tandon, ``The capacity of cache aided private information retrieval,'' in
  \emph{2017 55th Annual Allerton Conference on Communication, Control, and
  Computing (Allerton)}.\hskip 1em plus 0.5em minus 0.4em\relax IEEE, 2017, pp.
  1078--1082.

\bibitem{wei2018fundamental}
Y.-P. Wei, K.~Banawan, and S.~Ulukus, ``Fundamental limits of cache-aided
  private information retrieval with unknown and uncoded prefetching,''
  \emph{IEEE Transactions on Information Theory}, vol.~65, no.~5, pp.
  3215--3232, 2018.

\bibitem{8362308}
Y.~{Wei}, K.~{Banawan}, and S.~{Ulukus}, ``Private information retrieval with
  partially known private side information,'' in \emph{2018 52nd Annual
  Conference on Information Sciences and Systems (CISS)}, March 2018, pp. 1--6.

\bibitem{wan2019coded}
K.~Wan and G.~Caire, ``On coded caching with private demands,'' \emph{arXiv
  preprint arXiv:1908.10821}, 2019.

\bibitem{kamath2019demand}
S.~Kamath, ``Demand private coded caching,'' \emph{arXiv preprint
  arXiv:1909.03324}, 2019.

\bibitem{sarvepalli2019subpacketization}
V.~R. Aravind, P.~Sarvepalli, and A.~Thangaraj, ``Subpacketization in coded
  caching with demand privacy,'' \emph{arXiv preprint arXiv:1909.10471}, 2019.

\bibitem{wan2019device}
K.~Wan, H.~Sun, M.~Ji, D.~Tuninetti, and G.~Caire, ``Device-to-device private
  caching with trusted server,'' \emph{arXiv preprint arXiv:1909.12748}, 2019.

\bibitem{9330765}
Q.~{Yan} and D.~{Tuninetti}, ``Fundamental limits of caching for demand privacy
  against colluding users,'' \emph{IEEE Journal on Selected Areas in
  Information Theory}, pp. 1--1, 2021.

\bibitem{9174083}
X.~{Zhang}, K.~{Wan}, H.~{Sun}, and M.~{Ji}, ``Cache-aided multiuser private
  information retrieval,'' in \emph{2020 IEEE International Symposium on
  Information Theory (ISIT)}, 2020, pp. 1095--1100.

\bibitem{maddah2014fundamental}
M.~A. Maddah-Ali and U.~Niesen, ``Fundamental limits of caching,''
  \emph{Information Theory, IEEE Transactions on}, vol.~60, no.~5, pp.
  2856--2867, 2014.

\bibitem{9155254}
X.~{Zhang}, K.~{Wan}, H.~{Sun}, M.~{Ji}, and G.~{Caire}, ``Private cache-aided
  interference alignment for multiuser private information retrieval,'' in
  \emph{2020 18th International Symposium on Modeling and Optimization in
  Mobile, Ad Hoc, and Wireless Networks (WiOPT)}, 2020, pp. 1--8.

\bibitem{8226776}
Q.~{Yu}, M.~A. {Maddah-Ali}, and A.~S. {Avestimehr}, ``The exact rate-memory
  tradeoff for caching with uncoded prefetching,'' \emph{IEEE Transactions on
  Information Theory}, vol.~64, no.~2, pp. 1281--1296, Feb 2018.

\end{thebibliography}

\end{document}